\documentclass[conference]{IEEEtran}
\usepackage{amsmath,amssymb,amsfonts}
\usepackage{algorithm,algpseudocode}
\usepackage{graphicx}
\usepackage{url}
\usepackage{xcolor}
\usepackage{eso-pic}

\AddToShipoutPicture*{\footnotesize\sffamily\raisebox{1cm}{\hspace{1.65cm}\fbox{\parbox{\textwidth}{\copyright~2019 IEEE. Personal use of this material is permitted. Permission from IEEE must be obtained for all other uses, in any current or future media, including reprinting/republishing this material for advertising or promotional purposes, creating new collective works, for resale or redistribution to servers or lists, or reuse of any copyrighted component of this work in other works.}}}}

\newcommand{\trans}{\!\top}
\newcommand{\partialdndt}{\frac{\mathrm \partial}{\mathrm \partial t}}
\newcommand{\dndt}{\frac{\mathrm d}{\mathrm dt}}
\newcommand{\fitvec}   [1]{{\rm\bf #1}}
\newcommand{\fMeps}{{\bf M}_{\varepsilon}}
\newcommand{\fMkap}{{\bf M}_{\kappa}}
\newcommand{\fMsig}{{\bf M}_{\sigma}}
\newcommand{\fMnu}{{\bf M}_{\nu}}
\newcommand{\C}{\mbox{\rm\bf  C}}

\newcommand{\Gr} {\rm{\bf G}}

\newcommand{\fvphi}{\protect\boldsymbol{\rm\bf \varphi}}
\newcommand{\fitmat}[1]{{\mbox{\rm\bf #1}}}
\DeclareMathOperator{\divergence}{div}
\DeclareMathOperator{\curl}{curl}
\DeclareMathOperator{\grad}{grad}

\begin{document}

\title{A Darwin Time Domain Scheme for the Simulation of Transient
Quasistatic Electromagnetic Fields Including Resistive, Capacitive
and Inductive Effects}

\author{
\IEEEauthorblockN{1\textsuperscript{st} Markus Clemens}
\IEEEauthorblockA{\textit{University of Wuppertal} \\
\textit{Chair of Electromagnetic Theory}\\
Wuppertal, Germany \\
clemens@uni-wuppertal.de } \and
\IEEEauthorblockN{2\textsuperscript{nd} Bernhard K\"ahne}
\IEEEauthorblockA{\textit{University of Wuppertal} \\
\textit{Chair of Electromagnetic Theory}\\
Wuppertal, Germany\\
kaehne@uni-wuppertal.de} \and
\IEEEauthorblockN{3\textsuperscript{rd} Sebastian Sch\"ops}
\IEEEauthorblockA{\textit{Technische Universit\"at Darmstadt} \\
\textit{Graduate School of Computational Engineering}\\
Darmstadt, Germany \\
schoeps@gsc.tu-darmstadt.de}
}

\maketitle

\begin{abstract}
The Darwin field model addresses an approximation to Maxwell's
equations where radiation effects are neglected. It allows to
describe general quasistatic electromagnetic field phenomena
including inductive, resistive and capacitive effects. A Darwin
formulation based on the Darwin-Amp{\`e}re equation and the
implicitly included Darwin-continuity equation yields a
non-symmetric and ill-conditioned algebraic systems of equations
received from applying a geometric spatial discretization scheme and
the implicit backward differentiation time integration method. A
two-step solution scheme is presented where the underlying
block-Gauss-Seidel method is shown to change the initially chosen
gauge condition and the resulting scheme only requires to solve a
weakly coupled electro-quasistatic and a magneto-quasistatic
discrete field formulation consecutively in each time step. Results
of numerical test problems validate the chosen approach.
\end{abstract}


\section{Introduction}
\IEEEPARstart{Q}{uasistatic} field models derived from Maxwell's
equations are considered valid, if the shortest wavelength of a
problem well exceeds the diameter of the considered problem
\cite{SteinmetzKurzClemens2011:01},
\cite{MazauricRondotWendling2013:01s}. In these cases, radiation
effects can be neglected. For further taxonomy, the electric energy
density $w_e$ and magnetic energy density $ w_m$ are considered: in
case of $w_e \gg w_m$ everywhere in the problem domain, the
electro-quasistatic model is applicable and a variation of the
magnetic electric can be neglected, i.e., $\frac{\partial}{\partial
t} \bf{B} \approx 0.$ Thus, the electric field is irrotational
$(\curl \bf{E}=0)$ and governed by resistive and capacitive effects.
In case of $w_e \ll w_m,$ the magneto-quasistatic field
approximation only takes into account resistive and inductive field
effects where displacement currents are neglected within Amp\`ere's
law, i.e., $\frac{\partial}{\partial t} \bf{D} \approx 0.$

Quasistatic field scenarios, where $w_e \approx w_m$ holds, i.e.,
where capacitive, resistive and inductive field effects are to be
considered in the same problem, are often described using the full
set Maxwell's equations. As a result, in these models the otherwise
negligible radiation effects are still considered as an unnecessary
part of the model. Especially in time domain formulations this
results in a high stiffness of the resulting discrete field
formulations. Alternatively, quasistatic models of such scenarios
often involve the use of lumped parameter formulations based on
Kirchhoff's equations.

The Darwin field model is an approximation to Maxwell's equations
related to general quasistatic field scenarios including capacitive,
resistive and inductive field effects, i.e., where only radiation
effects can be neglected \cite{RaviartSonnendrucker1996:01s,
Larsson2007:01s,LiaoYing2008:01s, inpFangLiaoYing2009:01s,
KochSchneiderWeiland2012:01s,Garcia2018Chapter1S,
inpBadicsetal2018:01s}.

Following the notation in \cite{Garcia2018Chapter1S}, in the
quasistatic Darwin field model, the electric field $\bf E$ is
subject to a decomposition

\begin{equation}
\bf{E}=\bf{E}_{\mbox{\scriptsize irr.}} + \bf{E}_{\mbox{\scriptsize
rem.}}, \label{E-Field_Decomposition}
\end{equation}

and split up into an {\it irrotational} part
$\bf{E}_{\mbox{\scriptsize irr.}}$ with $\curl
(\bf{E}_{\mbox{\scriptsize irr.}})= 0,$ for which a scalar electric
potential representation $\bf{E}_{\mbox{\scriptsize irr.}}=-\grad
(\varphi)$ exists, and a {\it remainder} part
$\bf{E}_{\mbox{\scriptsize rem.}}$  In
(\ref{E-Field_Decomposition}), $\divergence
(\bf{E}_{\mbox{\scriptsize rem.}})=0$ only holds in the uniquely
special case of a Helmholtz decomposition that is typically only
admissible for homogeneous material distributions. Alternatively,
$\divergence(\bf{E}_{\mbox{\scriptsize rem.}}$) may be nonzero, which
needs to be taken into account within an additional gauge.
Within the Darwin field model, the rotational parts of the
displacement current densities are neglected with $\frac{\partial
}{\partial t}(\varepsilon \bf{E}_{\mbox{\scriptsize rot}})= 0.$ This
essentially eliminates the hyperbolic character of the full
Maxwell's equations which is responsible for modeling wave
propagation phenomena and translates the Darwin model into a system
with only first order time derivatives. Based on these equations
several reformulations in terms of the magnetic vector potential
$\bf{A}$ and scalar potential $\varphi$ can be formulated.

Following this introduction, in section II a quasistatic field model
for time domain problems is formulated featuring the
Darwin-Amp\`ere's equation and its corresponding Darwin-continuity
equation in terms of electrodynamic potentials. Section III
describes the space and time discretzation of this Darwin
formulation resulting in ill-conditioned and non-symmetric
monolithic algebraic systems of equations. Section IV introduces a
two-step solution technique which requires only symmetric algebraic
systems to be solved. In Section V, numerical test results are
shown, including a discussion of the results, followed by a
conclusion.

\section{$(\bf{A},\varphi)$ Formulation for the Darwin Model}

The splitting of the electric field in (\ref{E-Field_Decomposition})
is expressed in terms of electrodynamic potentials, i.e., the
magnetic vector potential $\bf{A}$ and the scalar electric potential
$\varphi$, with $\bf{E}_{\mbox{\scriptsize rem.}}= -{\partialdndt}
{\bf A}$ and $\bf{E}_{\mbox{\scriptsize irr.}}= - \grad \varphi.$
Initially, the electric field then reads
\begin{equation}
    \bf{E} = -{\partialdndt} {\bf A} - \grad \varphi.
    \label{E-Field_Decomposition2}
\end{equation}

The Darwin formulation ignores the rotational parts of the
displacement current densities related to the radiation of
electromagnetic waves, i.e., in the electric displacement currents ,
i.e., $ \frac{\mathrm
\partial^2}{\mathrm \partial t^2} \bf{A} \approx 0.$

With this ansatz, the Amp\`ere's equation reduces to the
Darwin-Amp\`ere's equation
\begin{eqnarray}
   \curl (\nu \curl \bf{A}) \!
   +\! \kappa {\partialdndt} \bf{A}\!
   +\! \kappa  \grad \varphi\!
   +\! \varepsilon  \grad {\partialdndt}   \varphi \!
   = \bf{J}_{\mathrm{s}},
       \label{Analytic_Darwin-Ampere_A-phi}
\end{eqnarray}
where $\nu$ is the reluctivity, $\kappa$ the specific electric
conductivity, $\varepsilon$ the permittivity and $\bf{J}_{\mathrm{s}}$ denotes
a transient source current density.

For the solution of (\ref{Analytic_Darwin-Ampere_A-phi}) an
additional equation is required to describe the relation of the
magnetic vector potential $\bf A$ and the electric scalar potential
$\varphi.$ Due to the gauge invariance of the Darwin field model,
various Darwin formulations can be derived \cite{Larsson2007:01s,
Garcia2018Chapter1S,inpBadicsetal2018:01s}. If non-homogeneous
material distributions need to be considered in the field model, the
coupling of $\bf{A}$ and $\varphi$ in
(\ref{Analytic_Darwin-Ampere_A-phi}) by use of the implicitly
contained continuity equation appears as an obvious choice. Left
application of the divergence operator to the
Darwin-Amp\`ere equation (\ref{Analytic_Darwin-Ampere_A-phi}) yields
a Darwin-model continuity equation
\cite{KochSchneiderWeiland2012:01s}
\begin{eqnarray}
   \mbox{div}\!\left(\!\!
       \kappa \frac{\partial}{\partial \bf{t}}
       \bf{A}\!\!
   +\! \kappa       \grad \varphi\!
   +\! \varepsilon  \grad \frac{\partial}{\partial t} \varphi\!\!
   \right)
   \!\!\!\!\!\!&=&\!\!\!\!\!
   \mbox{div} \bf{J}_{\mathrm{s}}.
       \label{Analytic_Darwin-continuity_eqn}
\end{eqnarray}
It should be noted, that (\ref{Analytic_Darwin-continuity_eqn})
is not identical to the
original continuity equation $\divergence \bf{J}+
{\partialdndt}\rho=0 $
related to the full Maxwell model, where
$\rho$ is the electric space charge. Including the Gau{\ss}' law,
the full Maxwell model continuity equation contains the expression
$\varepsilon \frac{\partial^2}{\partial t^2}\bf(A)$ related to the
rotational parts of the displacement currents. These are
specifically omitted within the Darwin model.

\section{A Discrete Darwin Model Formulation}

For the general case of non-homogeneous material distributions, a
spatial discretization to the equations
(\ref{Analytic_Darwin-Ampere_A-phi}) and
(\ref{Analytic_Darwin-continuity_eqn}) is considered following
\cite{KochSchneiderWeiland2012:01s}. The application of a mimetic
discretization scheme as e.g. the finite
integration technique (FIT)
\cite{Weiland77:01s,ibClemensWeiland2001:01s}, Whitney finite
 element method (WFEM) \cite{Nedelec80:01s} or the
cell method \cite{ibTonti2001:01} yields the systems of matrix
equations
\begin{eqnarray}
   \!\!
   \C^{\trans} \fMnu   \C       \fitvec{a}
   +           \fMkap  \dndt    \fitvec{a}
   +           \fMkap  \Gr           \fvphi
   +           \fMeps  \Gr   \dndt   \fvphi
   \!\!\!\!\!&=&\!\!\!\!\!\!
        \fitvec{j}_{\mathrm{s}},
        \label{FIT_Darwin1}
   \\
   \!\!
     \Gr^{\trans} \fMkap     \dndt  \fitvec{a}
   + \Gr^{\trans} \fMkap \Gr        \fvphi
   + \Gr^{\trans} \fMeps \Gr \dndt  \fvphi
   \!\!\!\!\!&=&\!\!\!\!\!\!
     \Gr^{\trans} \fitvec{j}_{\mathrm{s}},
        \label{FIT_Darwin2}
\end{eqnarray}
where $\fitvec{a}$ is the degrees of freedom (dof) vector related to the
magnetic vector potential, $\fvphi$ is the dof vector of electric nodal scalar
potentials, $\C$ is the discrete curl operator matrix, $\Gr$ and $\Gr^{\trans}$
are discrete gradient and (negative) divergence operator matrices. The matrices
$\fMnu, \fMkap, \fMeps$ are the (possibly nonlinear) discrete material matrices
of reluctivities, conductivities and permittivities, respectively, and the
construction of these discrete Hodge operators depends on the specific
discretization scheme.

The discrete Darwin equations (\ref{FIT_Darwin1}) and (\ref{FIT_Darwin2}) can
be rewritten as a first order differential-algebraic system of equations
\begin{eqnarray}
\left[
   \begin{array}{cc}
   \fMkap              &              \fMeps \Gr\\
   \Gr^{\trans} \fMkap & \Gr^{\trans} \fMeps \Gr
   \end{array}
   \right]
    \dndt
    \left[
   \begin{array}{c}
   \fitvec{a}\\
   \fvphi
   \end{array}
   \right]
\nonumber\\
+
 \left[
   \begin{array}{cc}
   \C^{\trans} \fMnu \C  &              \fMkap \Gr\\
   \fitmat{0}            & \Gr^{\trans} \fMkap \Gr
   \end{array}
   \right]
    \left[
   \begin{array}{c}
   \fitvec{a}\\
   \fvphi
   \end{array}
   \right]
    =
   \left[
   \begin{array}{c}
   \fitvec{j}_s\\
   \Gr^{\trans} \fitvec{j}_s
   \end{array}
   \right].
    \label{First_order_system}
\end{eqnarray}

While the non-symmetry of the block matrices in
(\ref{First_order_system}) can be partially eliminated, the presence
of metallic objects in the problem domain will result in large
differences in the order of magnitude of the entries.

A time discrete Darwin model based on the monolithic time domain
formulation (\ref{First_order_system}) results from the application
of e.g. a first order convergent Euler backward differentiation
(BDF1) formula \cite{KochWeiland2011:01s} with a unconditionally
stable time step $\Delta t := t^{n+1}-t^{n}.$ With
$\fMsig:=\fMkap + \frac{1}{\Delta t}\fMeps$ the implicit time
stepping scheme requires to solve the algebraic system of equations
\begin{eqnarray}
   \left[
   \begin{array}{cc}
   \C^{\trans} \fMnu \C  + \frac{1}{\Delta t}\fMkap    &              \fMsig \Gr\\
   \frac{1}{\Delta t} \Gr^{\trans} \fMkap              & \Gr^{\trans} \fMsig \Gr
   \end{array}
   \right]
  \left[
   \begin{array}{c}
   \fitvec{a}\\
   \fvphi
   \end{array}
   \right]^{n+1}\nonumber\\
   \!\!=\!\!\frac{1}{\Delta t}
 \left[
   \begin{array}{cc}
   \fMkap                            &              \fMeps \Gr\\
   \Gr^{\trans} \fMkap               & \Gr^{\trans} \fMeps \Gr
   \end{array}
   \right]
  \left[
   \begin{array}{c}
   \fitvec{a}\\
   \fvphi
   \end{array}
   \right]^{n}
   +
  \left[
   \begin{array}{c}
   \fitvec{j}_{\mathrm{s}}\\
   \Gr^{\trans} \fitvec{j}_{\mathrm{s}}
   \end{array}
   \right]^{n+1}
        \label{FIT_Darwin_Monolithic_BDF1}
\end{eqnarray}
for each time step and Newton iteration in case on nonlinear material laws.
The real-valued system matrix is non-symmetric
and can not be symmetrized; it is singular, if $\fMkap$ is singular.
In \cite{KochWeiland2011:01s} the system
(\ref{FIT_Darwin_Monolithic_BDF1}) is additionally regularized
assuming a non-physical conductivity in the non-conductive regions
of the simulation problem. The system matrices of the monolithic
time discrete algebraic systems of equations
(\ref{FIT_Darwin_Monolithic_BDF1}) can be extremely ill-conditioned
due to their off-diagonal matrix blocks with entries varying by
different orders of magnitude. In \cite{KochWeiland2011:01s} a
direct solver based on sparse LU-decomposition was used for a discrete
problem with a total number of only 12,101 dofs.

\section{A Two-Step Darwin Time Domain Scheme}
Rewriting the mutually coupled equations (\ref{FIT_Darwin1}) and
(\ref{FIT_Darwin2}) into
\begin{eqnarray}
   \!\!\!\!\!\!\!\!
   \C^{\trans} \fMnu   \C                  \fitvec{a}
   \!+\!
               \fMkap  \dndt               \fitvec{a}
   \!\!\!\!\!&=&\!\!\!\!\!
        \fitvec{j}_{\mathrm{s}}
   \!-\!           \fMkap  \Gr             \fvphi
   \!-\!           \fMeps  \Gr \dndt       \fvphi,
        \label{FIT_Darwin1a}
   \\
   \!\!\!\!\!\!\!\!
     \Gr^{\trans} \fMkap \Gr               \fvphi
   \!+\!
     \Gr^{\trans} \fMeps \Gr \dndt         \fvphi
   \!\!\!\!\!&=&\!\!\!\!\!
     \Gr^{\trans} \fitvec{j}_{\mathrm{s}}
   \!-\!
     \Gr^{\trans} \fMkap \dndt             \fitvec{a}
        \label{FIT_Darwin2a}
\end{eqnarray}
shows both the discrete magneto-quasistatic ($\bf{A}^{\star}$)
formulation (\ref{FIT_Darwin1a}) (see also
\cite{EmsonSimkin83:01s,Kameari90:02s,ClemensWeiland99:01s}) and the
discrete electro-quasistatic scalar electric potential formulation
(\ref{FIT_Darwin2a}) (see
\cite{ClemensWilkeBenderskayaDeGersemKochWeiland2004:01s}) coupled
to each other with their specific right hand side vectors,
respectively.

This motivates adopting a strong coupled iteration approach for the
solution of the time and space discretized reformulations of
(\ref{FIT_Darwin1a}) and (\ref{FIT_Darwin2a}). This approach is
identical to an iterative block-Gauss-Seidel solution of the
monolithic system (\ref{FIT_Darwin_Monolithic_BDF1}) whose convergence
can be shown by inspecting the spectral properties of the matrices.

Let's denote by $\fitvec{a}_{m}^{n+1}$ and $\fvphi_{m}^{n+1}$ the iterative
solution after the $m$-th Gauss-Seidel iteration at time step $n+1$.
The iteration scheme requires an initial guess, e.g. by extrapolation
\begin{equation}
    \label{extrap}
    \fitvec{a}_0^{n+1}:= \fitvec{a}^{n},
\end{equation}
where $\fitvec{a}^{n}$ denotes the final solution at time step $n$, i.e.,
after $m=1,2,\ldots$ iterations.

Due to discrete
conservation properties of the discrete field formulations
\cite{ibClemensWeiland2001:01s}, the discrete divergence relation
\begin{equation}
    \Gr^{\trans}\fMkap\fitvec{a}_{m+1}^{n+1}
    =
    \Gr^{\trans}\fMkap\fitvec{a}_{m}^{n+1}
    \label{FIT_DiscreteDivergence}
\end{equation}
holds. When using (\ref{extrap}), this discrete
conservation property (\ref{FIT_DiscreteDivergence}) implicitly
eliminates the time discrete backward derivative expression
$\frac{1}{\Delta t} \Gr^{\trans} \fMkap \left[\fitvec{a}_{m+1}^{n+1}
-\fitvec{a}^{n} \right]$ within the iteration scheme. This translates
to $\mbox{div}\!\left(\! \kappa \frac{\partial}{\partial \bf{t}}
\bf{A}\!\right)=0$ in the continuous case, i.e., the
rotational parts of the eddy current densities will be solenoidal.
This is an physically acceptable model assumption in case of highly
conductive materials within which the effects of the displacement
currents are negligible. Due to this, the Darwin continuity equation
(\ref{Analytic_Darwin-continuity_eqn})
reduces to the standard electro-quasistatic field formulation
\cite{ClemensWilkeBenderskayaDeGersemKochWeiland2004:01s}.

This property weakens the coupling of (\ref{FIT_Darwin1a}) and
(\ref{FIT_Darwin2a}) and the iterative scheme reduces to the two-step
approach shown in Algorithm \ref{alg:Two-Step_Darwin_TD_BDF}, where
in each time step only two symmetric and positive (semi-)definite
systems of algebraic equations need to be solved consecutively. For
this task efficient solution schemes are available (see e.g.
\cite{ClemensWeiland99:01s,
ClemensWilkeBenderskayaDeGersemKochWeiland2004:01s,
petsc-efficient}).

\begin{algorithm}[h]
    \caption{Two-Step Darwin Time Domain Algorithm}
    \label{alg:Two-Step_Darwin_TD_BDF}
    \begin{algorithmic}[1]
        \State Initialize $\fvphi^{0}:=\fvphi (t_0);
        \fitvec{a}^{0}:=\fitvec{a}(t_0);$
       \For{$n\gets 0:n_{\mbox{\scriptsize End}}$:}
            \State Solve:
            \State $\left[ \Gr^{\trans} \fMsig \Gr \right]
                    \fvphi^{n+1}
                    =
                    {\Gr}^{\trans}\! \fitvec{j}_s^{n+1}
                    \!\!+ \!
                    \frac{1}{\Delta t}{\Gr}^{\trans} \!\fMeps {\Gr}\fvphi^{n}
                    ;$
            \State Solve:
            \State $\bigl[\C^{\trans} \fMnu \C + \frac{1}{\Delta t} \fMkap \bigr] \fitvec{a}^{n+1}$
            \State $=
                    \fitvec{j}_s^{n+1}
                    +
                    \frac{1}{\Delta t} \fMkap \fitvec{a}^{n}
                    -
                    \fMsig {\Gr} \fvphi^{n+1}
                    +
                    \frac{1}{\Delta t} \fMeps   {\Gr} \fvphi^{n};$
       \EndFor
    \end{algorithmic}
\end{algorithm}

\section{Numerical Results}
Two test structures are excited at $f=100$ MHz, with
dimensions small against the wave length, i.e., the quasistatic
assumption holds. Using ramped sinusoidal excitations, the two-step
Darwin time domain scheme is implemented using the MFEM library
\cite{misc_MFEM:01}. For the solution of the electro-quasistatic
system and the weakly coupled magneto-quasistatic systems at each
time step, efficient algebraic multigrid (AMG) schemes provided in
the PETSc \cite{petsc-efficient} linear algebra solver library are
used.

\subsection{High-frequency coil}
\begin{figure}[htbp]
   \centering
   \includegraphics[width=0.45\textwidth]{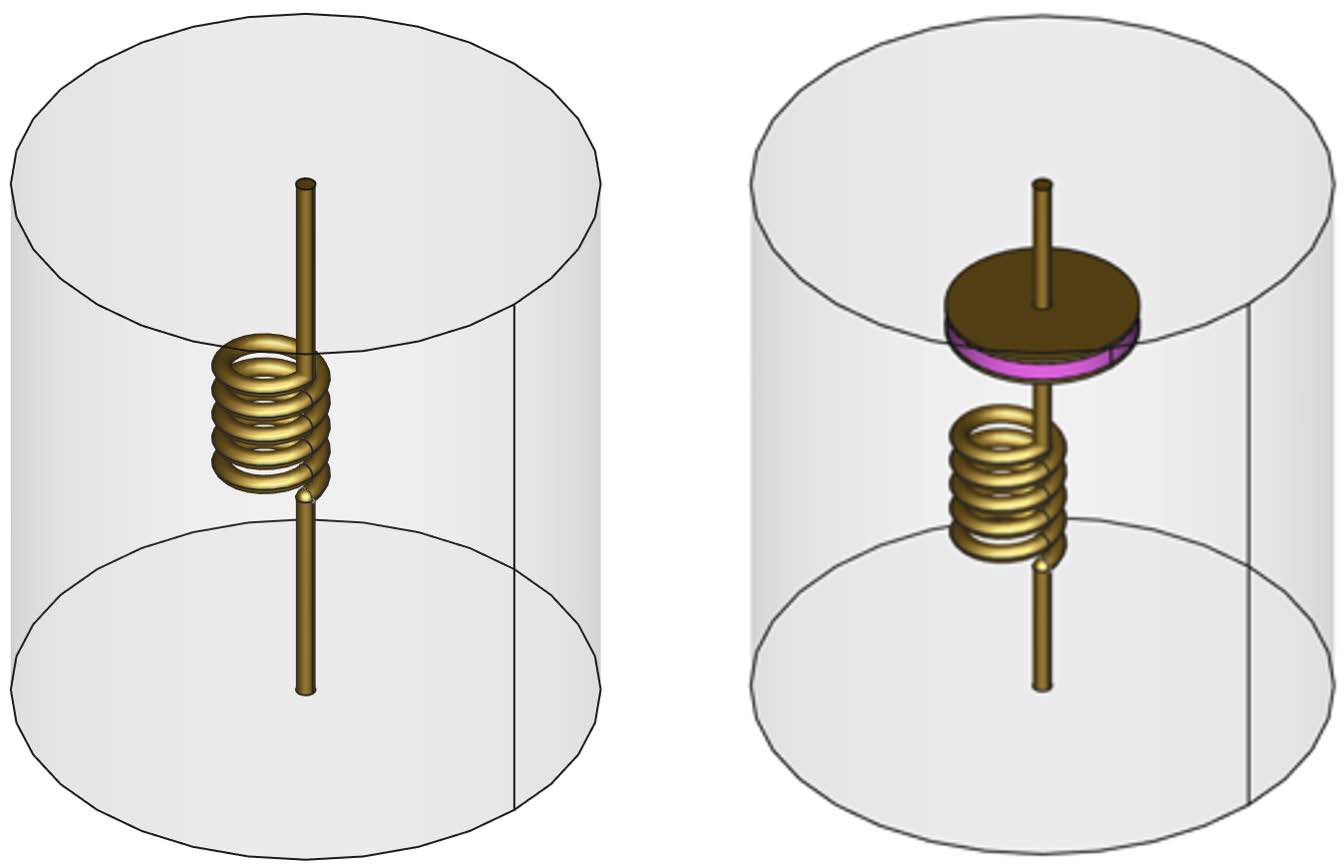}
   \caption{Coil geometry (left); RLC structure geometry (right).}
   \label{fig:1}
\end{figure}

\begin{figure}[htbp]
   \centering
   \includegraphics[scale=0.39]{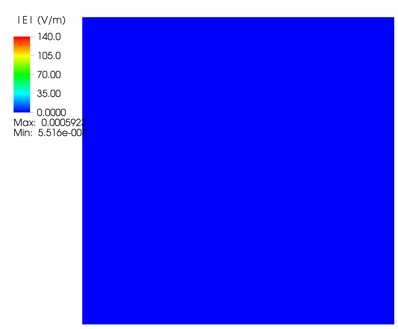}
   \includegraphics[scale=0.39]{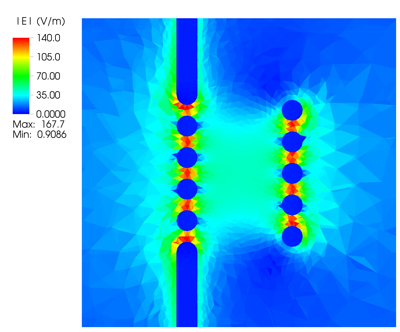}
   \includegraphics[scale=0.39]{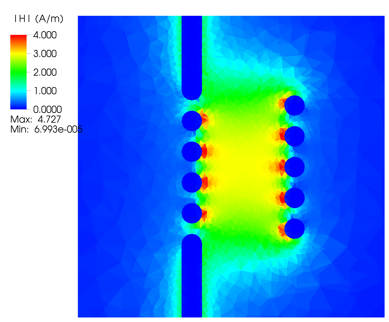}
   \includegraphics[scale=0.39]{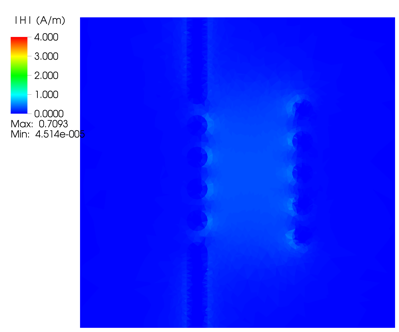}
   \caption{Magnitudes of the electric field $E(t=9T)$ (top left)
   and magnetic field $B(t=9T)$ (bottom left) and
   the electric field $E(t=9,25 T)$ (top right) and magnetic field $B(t=9.25 T)$ (bottom right)
   simulated with the two-step Darwin time domain algorithm.}
   \label{fig:2}
\end{figure}

A high-frequency coil structure of 63 mm length (see Fig.
\ref{fig:1} (left) is considered using a mesh consisting of 237,835
tetrahedra. For the time-domain simulation a ramped sinusoidal
excitation profile is used for 10 periods. The dof vector $\fvphi$
has dimension 39,038 and the vector $\fitvec{a}$ has dimension
$278,090$. Fig \ref{fig:2} shows results achieved for the magnetic
and the electric field after $t=9 1/f$ and $t=9,25 1/f$.

\subsection{RLC Model}
\begin{figure}[htbp]
   \centering
   \includegraphics[scale=0.39]{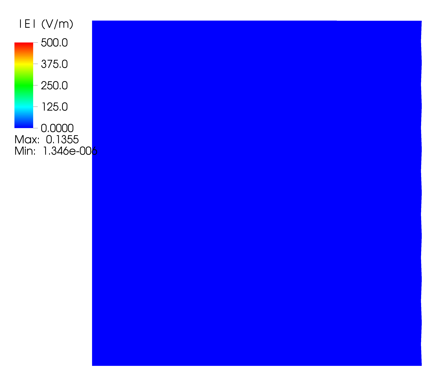}
   \includegraphics[scale=0.39]{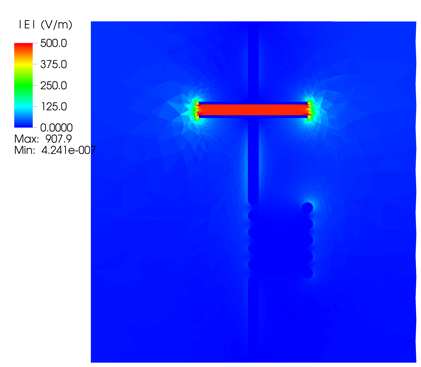}
   \includegraphics[scale=0.39]{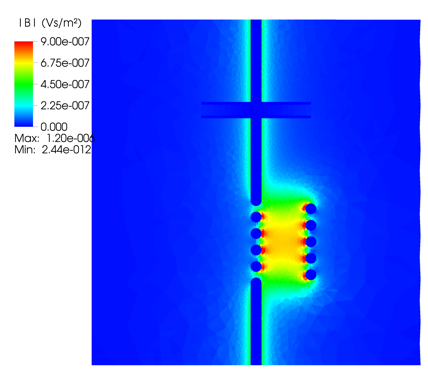}
   \includegraphics[scale=0.39]{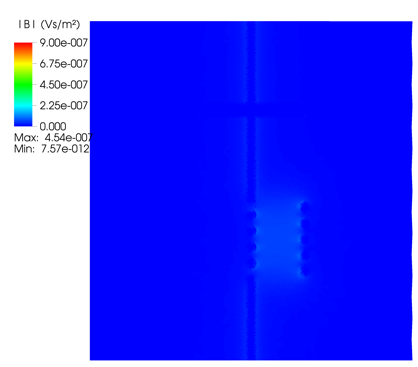}
   \caption{Magnitudes of the electric field $E(t=9T)$ (top left)
   and magnetic field $B(t=9T)$ (bottom left) and
   the electric field $E(t=9,25 T)$ (top right) and magnetic field $B(t=9.25 T)$ (bottom right)
   simulated with the two-step Darwin time domain algorithm.}
   \label{fig:4}
\end{figure}

A RLC-structure of 63 mm length presented in
\cite{JochumfarleDyczili-Edlinger2015:01s,inpBadicsetal2018:01s},
i.e., a wire (electrical conductivity $\kappa = 10^6$\;S/m
connecting a coil and a capacitor with a dielectric inset
($\varepsilon_{r} = 2$) (see Fig. \ref{fig:1} (right)) is
considered. The FEM mesh consisting of 543,783 tetrahedra. For the
time-domain simulation a ramped sinusoidal excitation profile is
used for 10 periods. The dof vector $\fvphi$ has dimension $88,273$
and the vector $\fitvec{a}$ has dimension $633,542 $.
Fig \ref{fig:4} shows the simulation results achieved with the
two-step Darwin time domain scheme for the magnetic and the electric
field.

\subsection{Discussion}
The simulation results achieved with the two-step time domain
method, show that resistive, inductive and capacitive effects are
included: The capacitive coupling of the high-frequency coil
windings in the HF coil is included with the irrotational parts of
the electric field and shown in Fig. \ref{fig:2}. The results in
Fig. \ref{fig:2} and Fig. \ref{fig:4} show the exchange of electric
and magnetic field energy in the sinusoidally excited test
structures.

\section{Conclusion}
The Darwin field model was analyzed to describe general quasistatic
electric and magnetic field distributions by only neglecting the
rotational contributions of the displacement currents. Starting from
an ($\bf{A},\varphi$) formulation of the Darwin-Amp\`ere law and the
Darwin-continuity equation, the resulting discrete Darwin model
represents a differential-algebraic set of equations. In order to
avoid the solution of ill-conditioned and non-symmetric monolithic
algebraic systems of equations required within implicit time
discretization schemes, a two-step solution schemes was presented
based on the consecutive solution of weakly coupled discrete
electro- and magneto-quasistatic field formulations in each time
step. Numerical results of quasistatic electromagnetic structures,
where capacitive, inductive and resistive field effects need to be
considered, showed the validity of this two-step approach and its
ability to solve realistic 3D problem resolutions by enabling the
use of efficient solution schemes.

\section*{Acknowledgement}
This work is supported in parts by the Deutsche Forschungsgemeinschaft (DFG) under grant no. CLE143/10-2.

\vspace{0,5cm}

\end{document}